\documentclass[prl,twocolumn,showpacs,preprintnumbers,superscriptaddress]{revtex4}

\usepackage{newlfont}
\usepackage{amssymb}
\usepackage{amsfonts}
\usepackage{amsmath}
\usepackage{amsthm}
\usepackage{graphicx}
\usepackage{epsfig}
\usepackage{color}
\usepackage{bm}

\newcommand{\ket}[1]{|{#1}\rangle}
\newcommand{\bra}[1]{\langle{#1}|}
\newcommand{\braket}[2]{\langle {#1} | {#2} \rangle}
\renewcommand{\t}[1]{\textrm{#1}}
\newcommand{\com}[1]{[\textsc{#1}]}

\begin{document}
\title{Physical Cost of Erasing Quantum Correlation}

\author{Arun Kumar Pati}

\affiliation{Quantum Information and Computation Group,\\
Harish-Chandra Research Institute, Chhatnag Road, Jhunsi, 
Allahabad 211 019, India}


\begin{abstract}
Erasure of information stored in a quantum state requires energy cost and is inherently an irreversible operation.
If quantumness of a system is physical, does erasure of quantum correlation as measured by discord also need some 
energy cost? Here, we show that change in quantum correlation is never larger than the total entropy change of the system and
the environment. The entropy cost of erasing correlation has to be at least equal to the amount of quantum correlation erased. 
Hence, quantum correlation can be regarded as genuinely physical.
We show that the new bound leads to the Landauer erasure. The physical cost of erasing quantum correlation is well 
respected in the case of bleaching of quantum information, 
thermalization, and can have potential application for any channel leading to erasure of quantum correlation.

\end{abstract}

\pacs{03.67.Hk, 05.40.Ca}

\maketitle

The notion that ``information is physical''-- has lead to profound insights about the nature of information stored in classical and quantum systems. This is the essence of 
the Landauer principle which states that the erasure of information requires some energy cost and is necessarily an irreversible operation \cite{sz,land}.
This has played an important role in the emerging field of quantum information science as storing and processing of information require
physical systems and physical laws \cite{ben,zurek,lub,seth,martin,ved,rend}. Thus, quantum information stored in a qubit is physical. Similarly, the notion of entanglement which is considered
as a very useful resource is also physical. To store and manipulate entanglement we need composite quantum systems and quantum mechanical allowed 
operations. 
Using quantum entanglement one can perform informational theoretic  tasks such as quantum teleportation \cite{tele}, 
dense coding \cite{dc}, 
remote state preparation \cite{remote,ben1} and many more. Can quantum entanglement also do physical work? In fact, recently it has been shown that the 
cost of erasure of information in the presence of quantum memory can depend on the conditional entropy of the system and the memory \cite{ren}. If there 
is entanglement present between them, then the erasure of information can gain work instead of spending it.

In recent years, there have been several efforts in trying to understand the nature of correlations present in the composite quantum systems 
\cite{hend,zurek1,jon,raja,zurek2,grois,sen,bruk,modi}. In particular, 
there can be a situation where there is no entanglement yet one expects some quantumness \cite{knill,animesh}. A promising candidate for this is so called
quantum discord which captures the quantum correlation of any bipartite state \cite{zurek1}. 
Much of the current work are devoted in trying to find operational meanings for quantum discord and its usefulness in quantum 
communication \cite{madhok,caval,alex,chuan,modi1}. In one of the early attempts, it was shown that 
the quantum discord is related to the extractable work difference by a quantum and a classical demon \cite{zurek2}.
Here, we raise the
question: can quantum correlation be considered as a real physical quantity like information? If it is so, does erasing quantumness of a 
bipartite state necessarily involve some energy cost? Erasure of quantum correlation would mean that we start with a bipartite 
quantum correlated state and by allowing one of the subsystem to interact with an environment the bipartite quantum state transforms to a 
zero discord state. This may be called as the complete erasure of quantum correlation under local operation. In this paper, we show that quantum discord is 
not just an informational resource but it is a 
physical one. We prove that whenever there is a change in quantum correlation that should be accompanied by entropy generation in the system and 
the environment. 
We also show that the Landauer erasure can be obtained from our entropic bound.
Further, we show that any positive creation of quantum correlation must be at least equal to the negentropy of 
the combined system and the environment. Our results provide new insights to the physical nature of quantum correlation irrespective of any particular 
quantum task or quantum communication protocol.

Let $\rho_{AB}$ is a density operator that describes a bipartite system $AB$. It is a positive operator that acts on the Hilbert space ${\cal H} =
{\cal H_A} \otimes {\cal H_B}$. The reduced states $\rho_A = \mbox{Tr}_B \left[ \rho_{AB} \right]$ and 
$\rho_B = \mbox{Tr}_A \left[ \rho_{AB} \right]$ represent the state of the subsystem $A$ and $B$ and they act on ${\cal H}_A$ and 
${\cal H}_B$, respectively. Quantum discord for a bipartite quantum state $\rho_{AB}$ is the difference between two equivalent definitions of mutual information.
The quantum mutual information of any bipartite state is obtained from 
the classical definition of mutual information 
by replacing the Shannon entropies by the von Neumann entropies. For a bipartite quantum state \(\rho_{AB}\), it  
is defined as \cite{cerf}
\begin{equation}
\label{qm}
I(\rho_{AB})= S(\rho_A)+ S(\rho_B)- S(\rho_{AB}),
\end{equation}
where \(S(\sigma) = - \mbox{Tr} \left(\sigma \log_2 \sigma\right)\) is the von Neumann entropy of a quantum state \(\sigma\). 
The quantum mutual information is a measure of total correlation (classical and quantum) for any bipartite state. An operational 
meaning for quantum mutual information is that it is the minimal amount of randomness that is required to completely erase all the 
correlations in $\rho_{AB}$ \cite{grois}.

If an observer is trying to extract information by doing measurement on $B$, then 
the classical correlation is given by  \cite{hend}
\begin{equation}
\label{cmi1}
J_B(\rho_{AB}) = S(\rho_A) - \min_{\{\Pi_i^B\}} \sum_i p_i S(\rho_{A|i}),
\end{equation}
where the conditional entropy upon measurement is $S(\rho_{A|B}) = \min_{\{\Pi_i^B\}} \sum_i p_i S(\rho_{A|i})$.
The minimization being over all POVMs performed on subsystem \(B\).
The probability for obtaining \(i\) outcome is \(p_i = \mbox{Tr}_{AB}[(I_A \otimes \Pi^B_i) \rho_{AB} (I_A \otimes \Pi^B_i) ]\), and 
the corresponding post-measurement state  
for the subsystem \(A\) is given by \(\rho_{A|i} = \frac{1}{p_i} \mbox{Tr}_B[(I_A \otimes \Pi^B_i) \rho_{AB} (I_A \otimes \Pi^B_i)]\), 
where \( {I}_A\) is the identity operator on \( {\cal H}_A\). 
The quantity $J_B(\rho_{AB})$ is the classical correlation and can be interpreted as information gain about subsystem $A$ as a result of a
measurement on the subsystem $B$. The difference $[I(\rho_{AB})-J_B(\rho_{AB})]$ is a measure of quantum correlations, and called as the quantum discord. 
Since the quantum mutual information is never lower than the classical correlation, discord is always positive. Therefore, the 
quantum discord with respect to measurement on $B$ is given by 
\cite{zurek1}
\begin{equation}
D_B(\rho_{AB})= 
\min_{\{\Pi_i^B\}} \sum_i p_i S(\rho_{A|i})  - S(A|B),
\end{equation}
where $S(A|B) = S(\rho_{AB}) - S(\rho_{B})$ is the conditional entropy. 
It is known that unlike quantum entanglement, even some separable states have a nonzero discord. 

Here, we ask the question: what is the thermodynamic cost of changing the quantum correlation when we perform general quantum operation on one of the subsystem 
(say $B$). When $\rho_{AB} \rightarrow \rho_{AB}' = (I_A \otimes {\cal E}_B )( \rho_{AB})$, the amount of change in the quantum correlation is ~~
$\Delta D(A:B) = D_B(\rho_{AB}) - D_B(\rho_{AB}')$. If the final state $\rho_{AB}' = \sum_i p_i \mbox{Tr}_B [{\cal E}_B(\rho_{AB})] \otimes 
|i\rangle \langle i |$ is a quantum-classical state, then this we call as the 
complete erasure of quantum correlation under local operation. Since any quantum operation 
can be represented as a unitary evolution in an enlarged Hilbert space, the erasure operation is given by 
$\rho_{AB} \otimes \rho_E \rightarrow \rho_{ABE}' = 
(I_A \otimes U_{BE})(\rho_{AB} \otimes \rho_{E}) (I_A \otimes U_{BE})
^{\dagger}$, where $\rho_E$ is the initial state of the ancilla (environment).
When a quantum system interacts with the environment, both of them can get correlated. As a result of this interaction, the original information
about the quantum state may be lost. However, as proved by the no-hiding theorem \cite{sam} in quantum theory if information is completely bleached out from the 
original state, then it moves to some subspace of the environment Hilbert space with no information stored in the bipartite correlation. If the 
the original system itself is a composite object and information is bleached out, then the quantum correlation between the original
system can change as it gets more correlated with the environment.
Now, we quantify the amount of change in the quantum correlation when the quantum system in a state $\rho_{AB}$ interacts with an external world (say) 
environment in a state $\rho_E$.

{\bf Theorem} 
Let $\rho_{AB}$ be a state on ${\cal B}({\cal H}_A \otimes {\cal H}_B)$ and ${\cal E}_B$ a trace preserving  completely positive 
map acting on the subsystem $B$ with $\rho_{AB} \rightarrow \rho_{AB}' = 
(I_A \otimes {\cal E}_B )( \rho_{AB})$. 
The amount of change in the quantum correlation $\Delta D(A:B) = D_B(\rho_{AB}) - D_B(\rho_{AB}')$
can never be larger than the total entropy change of the system and the environment, i.e.,~~ $\Delta D(A:B) \le \Delta S_{T}$.

{\em Proof:}  Using the definition of the quantum discord, the change in the quantum correlation can be expressed as
\begin{eqnarray}
\Delta D(A:B) &=& [S'(A|B) - S(A|B)] \nonumber\\
&-& [\min_{\{\Pi_i^B\}} \sum_i p_i S(\rho_{A|i}) - \min_{\{\Pi_i^B\}} \sum_i p_i' S(\rho_{A|i}')]. \nonumber\\
\end{eqnarray}
Now we can write the above as
\begin{eqnarray}
\Delta D(A:B) &=& \Delta S_{T} - [J(\rho_{AB}) - J(\rho_{AB})'] \nonumber\\
&+& [S(\rho_{B}) + S(\rho_{E})] - [S(\rho_{B}') + S(\rho_{E}')],
\end{eqnarray}
where $\Delta S_{T} = \Delta S_{AB} + \Delta S_{E}$ is the total change in the entropy of the system $AB$ and the environment $E$.
We define the initial and final total entropies as sum of the entropies of the bipartite quantum system $AB$ and the environment $E$ before and after 
the interaction, respectively as
\begin{eqnarray}
\label{ent}
S_{T} &=& S(\rho_{AB}) + S(\rho_{E}), \nonumber\\
S_{T}' &=& S(\rho_{AB}') + S(\rho_{E}'),
\end{eqnarray}
where $\rho_{AB}' = \mbox{Tr}_E \left[ \rho_{ABE}' \right]$ and $\rho_{E}' = \mbox{Tr}_{AB} \left[ \rho_{ABE}' \right]$. Then the change in total entropy is 
defined as
$\Delta S_{T} = S_{T}' - S_{T} =  \Delta S_{AB} + \Delta S_{E}$.
It may be noted that the change in the total entropy is always positive which follows from the subaddivity of entropy
and the unitarity .

Now, note that $\Delta J(A:B) = [J_B(\rho_{AB}) - J_B(\rho_{AB}')]$ is the change in the classical correlation. 
As classical correlation cannot increase under local operation on Bob, i.e., when $\rho_{AB} \rightarrow {\cal E}_B (\rho_{AB})$, we have
$J_B(\rho_{AB}) \ge  J(\rho_{AB}')_B$. Since there is minimization over all POVMs involved in the classical correlation, 
the action of local operation on $B$ can be thought of as a part of the POVM on $B$ \cite{hend}. Thus, we have
\begin{eqnarray}
\Delta D(A:B) & \le &   \Delta S_{ T} + S(\rho_{BE}) - S(\rho_{BE'}).
\end{eqnarray}
Here, we use the fact that initially $B$ and $E$ are in product state, hence $ S(\rho_{B}) + S(\rho_{E}) =
 S(\rho_{BE})$ and using the subadditivity of the von Neumann entropy, we have $ S(\rho_{B}') + S(\rho_{E}') \ge 
 S(\rho_{BE}')$. Next,  we use $S(\rho_{BE} ) =  S(\rho_{BE}')$ as they evolve unitarily and hence we have
\begin{eqnarray}
\label{main}
\Delta D(A:B) & \le & \Delta S_{T}.
\end{eqnarray}
This shows that the amount of erased quantum correlation can never exceed that of the total entropy 
change. This shows that whenever there is change in quantum correlation there must be entropy generation. {\em The total entropy production 
in the system and the environment has to be at least equal to the change in the quantum correlation. } 

Suppose that the interaction of $B$ with the environment leads to a quantum measurement, in the sense that after the interaction the 
state of $AB$ is a quantum-classical state. In this case, we have 
$\rho_{AB}' = \mbox{Tr}_{E}[ (\mathbb{I}_A \otimes U_{BE}) (\rho_{AB}\otimes \rho_E ) (\mathbb{I}_A \otimes U_{BE})^{\dagger})] =
\sum_i p_i {\rho_i}_A \otimes |i\rangle_B \langle i|$. It is known that quantum discord is zero for such a state. This can be 
thought of as a physical process leading to complete erasure of quantum correlation. 
If we define entropy cost of erasing quantum correlation as $W(A:B)= k T \Delta S_T$, then we have 
\begin{eqnarray}
W(A:B) \ge k T ~ \Delta D(A:B) .
\end{eqnarray}
Thus, the erasure of $D(A:B)$-bits of quantum correlation has an energy cost.
The entropy cost of erasing correlation has to be at least equal to the amount of quantum correlation erased.  
If the initial state $\rho_{AB} = |\Psi\rangle_{AB} \langle \Psi|$ is pure, then the quantum discord is equal to the entanglement entropy 
$E(|\Psi \rangle_{AB}) =
S(\rho_A) = S(\rho_B)$. In this case, our result tells that the cost of erasing quantum entanglement is given by
$W(A:B) \ge  kT ~ S(\rho_A)$. This gives another thermodynamic meaning to entanglement entropy. It is the minimum amount of
energy (per $kT ~$) that is required to erase the quantumness shared between $A$ and $B$. 

Our result can also be applied to answer the question: what is the physical cost of creating quantum correlation? 
Since quantum discord is not monotonic, it might increase during the interaction with an 
environment. In fact, recent work shows that there are certain local dissipative channels which can create quantum discord \cite{dag,gio,cam}. 
In particular, if the initial state of $\rho_{AB}$ has no quantum correlation (quantum-classical state, with 
$D_B(\rho_{AB}) =0$), then after $B$ is subject to quantum operation ${\cal E}_B$, we may have $D_B(\rho_{AB}') > 0 $. Then our result tells that 
\begin{eqnarray}
 D_B(\rho_{AB}')  \ge - \Delta S_T.
\end{eqnarray}
This shows that the amount of quantum correlation created is at least equal to the total negentropy of the system and the environment. 
Thus, erasure and creation of quantum correlation both have thermodynamic cost.

Since discord is asymmetric with respect to measurement performed on the subsystem, one may ask what is the cost of erasing 
quantum correlation if we consider $D_A(\rho_{AB})$ instead of $D_B(\rho_{AB})$, where $D_A(\rho_{AB})$ is defined as
$D_A(\rho_{AB}) = I(\rho_{AB}) - J_A(\rho_{AB})$ and $J_A(\rho_{AB}) = S(\rho_B) - \min_{\{\Pi_i^A\}} \sum_i p_i S(\rho_{B|i})$.
The conditional entropy now is given by $S(\rho_{B|A}) = \min_{\{\Pi_i^A\}} \sum_i p_i S(\rho_{B|i})$ with 
the probability \(p_i = \mbox{Tr}_{AB}[( \Pi^A_i \otimes I_B) \rho_{AB} (\Pi^A_i \otimes  I_B )]\).
The conditional state for the subsystem \(B\) is \(\rho_{B|i} = \frac{1}{p_i} \mbox{Tr}_A [ (\Pi^A_i \otimes I_B) \rho_{AB} 
(\Pi^A_i \otimes I_B)] \), 
where \( {I}_B\) is the identity operator on \( {\cal H}_B\). Following theorem 1, we can show that when 
$\rho_{AB} \otimes \rho_E \rightarrow \rho_{ABE}' = ( {I}_A \otimes U_{BE})
(\rho_{AB} \otimes \rho_{E})
( {I}_A \otimes U_{BE})^{\dagger}$ the change in the quantum correlation $\Delta D(B:A) = D_A(\rho_{AB}) - D_A(\rho_{AB}')$ 
is given by $\Delta D(B:A) \le \Delta S_{ T}$. 
Here, $\Delta D(B:A)$ is always positive as it cannot increase under
quantum operation on unmeasured subsystem.

We can derive the Landauer erasure principle from our entropic bound (8). Let us consider the erasure process as  
a physical map that transforms $\rho_B \otimes \rho_E \rightarrow \rho_B' \otimes \rho_E'$, where $\rho_B$ and 
$\rho_E =\rho_{bath}$ are the initial states of the system and the bath, and similarly $'$ denotes the final states.
Let the bath is at a temperature $T$. Ideally, the erasure process transforms any arbitrary state $\rho_B$ to a fixed pure state (resetting).
One way to achieve this is by thermal randomization \cite{lub,martin}, where final state of the system may not be pure, 
but $\rho_B'$ can be arbitrarily close to a fixed pure state. Let the initial state of the bath is in a thermal state
$\rho_E = \exp(-\beta H_E)/Z$, where $H_E$ is the Hamiltonian of the bath and $Z$ is the partition function.
We can imagine that $\rho_B=  \mbox{Tr}[\rho_{AB}]$ for some density operator on ${\cal H}_{AB}$.
During the interaction of system and bath, the increase in the energy of the bath is $\Delta E_E = \Delta E 
= \mbox{Tr}[\rho_E' H_E] -
\mbox{Tr}[\rho_E H_E]$ and one can also see that the change in the entropy of system $B$ and $E$ satisfy 
$\Delta S_B + \Delta S_E \ge 0$ \cite{partovi}. Conservation of energy requires that $\Delta E_B + \Delta E_E =0$.
In particular, if the initial state of $AB$ is uncorrelated, i.e., $\rho_{AB} = \rho_A \otimes \rho_B$, then $\Delta D(A:B) =0$. Using
$\Delta S_{AB} \ge \Delta S_B$ and our bound (8) we have 
$\Delta E \ge k T S(\rho_B)$ which is precisely the well known Landauer's erasure principle.
If the system $B$ is correlated with another one $A$ (which may be called as quantum memory), then we have a generalized version 
of the Landauer erasure. Let us define the entropy cost of erasure as $\Delta S_{\mbox{\small{erasure}}} = \Delta S_B + \Delta S_E$ \cite{martin,ved}.  
Now using the fact that $\Delta S_T = \Delta S_{AB} + \Delta S_E \le S(\rho_B') - S(B|A) + \Delta S_E$, where $S(B|A) = 
S(\rho_{AB}) - S(\rho_A)$, we have
a new inequality 
\begin{eqnarray}
\label{new}
\Delta S_{\mbox{\small{erasure}}} \ge \Delta D(A:B) + S(B|A) - S(\rho_B).
\end{eqnarray}
Here we have not assumed that the final state $\rho_B'$ is pure (as in ideal erasure process).
Since the conditional entropy can be negative, this shows that the cost of erasing quantum information depends both on the quantum correlation and 
entanglement present between the system $B$ and another system $A$ which can act as a quantum memory. This can be interpreted as a generalization of 
Landauer's erasure in the presence of quantum memory. 

Our result can be applied to any physical scenario where there is loss of quantum correlation. We apply here 
to quantum bleaching and thermalization processes and show that indeed our bound is respected.
A quantum bleaching process is a perfect hiding process where an arbitrary quantum state $\rho$ is transformed 
to a fixed density operator $\sigma$ under a completely positive trace preserving (CPT) map and the final state has no information
whatsoever about the input state. 
At the same time, this can also be regarded as a generalization of Landauer's erasure operation (where an arbitrary
state $\rho$ is mapped to a pure state for resetting). It was proved that when information is bleached out from the original quantum system, it
moves to rest of the universe (environment or bath) with no information stored in the bipartite correlation \cite{sam}. 

Let us imagine that the 
system in a state $\rho$ undergoes a bleaching process.
The density matrix $\rho = \rho_B = \sum_j \lambda_j |b_j\rangle \langle b_j |$ can be purified to an entangled state 
$|\Psi\rangle_{AB} = \sum_j \sqrt{\lambda_j}|a_j \rangle_A |b_j \rangle_B $, with $\rho_B = \mbox{Tr}_A[|\Psi\rangle_{AB} \langle \Psi |]$ . 
Let the density matrix $\rho_E$ is a pure state with $\rho_E =  | E \rangle_{E} \langle E |$. The environment may be a composite system (for simplicity
we can imagine it as a bipartite system).
Under the hiding map, we have  $|\Psi\rangle_{AB} \langle \Psi | \otimes |E \rangle_{E} \langle E | \rightarrow 
\mathbb{I}_A \otimes U_{BE} |\Psi\rangle_{AB} \langle \Psi | \otimes |E \rangle_{E} \langle E | (\mathbb{I}_A \otimes U_{BE})^{\dagger}$
The pure state transformation for the quantum bleaching process is given by
\begin{eqnarray}
|\Psi\rangle_{AB} |E \rangle_{E} & \rightarrow & (\mathbb{I}_A \otimes U_{BE}) |\Psi\rangle_{AB} |E \rangle_{E}  \nonumber\\
& = & \sum_{jk} \sqrt{\lambda_j p_k}|a_j \rangle_A |k \rangle_B \otimes (|b_j \rangle |q_k \rangle)_E, \nonumber\\
\end{eqnarray}
where we have used the hiding map $|b_j \rangle_B | E \rangle_{E} \rightarrow
\sum_{k} \sqrt{p_k} |k \rangle_B \otimes (|b_j \rangle |q_k \rangle)_E $ (up to adding zero vectors in the environment Hilbert space) \cite{sam}.
The quantum bleaching process completely decorrelates the system $AB$.
In this process, the change in the quantum correlation is $D_B(\rho_{AB}) - D_B(\rho_{AB}') = S(\rho_B)$ as the initial state 
$\rho_{AB}$ is pure and the final state $\rho_{AB}' $ is a product state. 
Using various entropic quantities, one can check that our inequality is respected. Also, one can see that the amount of quantum correlation 
erased is compensated by the generation
of the same between the subsystem $A$ and one of the subspace of the environment Hilbert space.
This is nothing but a form of Landauer's principle which tells us that the amount of erased information can never be larger than
entropy of erasure. 
As information is now completely in the environment Hilbert space, the amount of quantum correlation
lost is exactly equal to the entropy of erasure.

Our result can be applied to thermalization of an arbitrary quantum state (not necessarily erasure). The process of thermalization is relaxation towards the 
thermal equilibrium of
a system in contact with a bath (environment). Let us consider the situation 
where our quantum system $\rho_B$ interacts with a bath in a thermal equilibrium state $\rho_E = \exp(-\beta H_E)/Z_E$, where 
$Z_E = \mbox{Tr}[\exp(-\beta H_E)]$ is the partition function. The system reaches equilibrium if its time evolved state approaches some equilibrium state $\rho_B' =
\rho_{eq}$ and spends 
most of the time close to it. Let us imagine again a purification of the initial state of $B$ and $E$. 
Let the density matrix $ \rho_B = \sum_n w_n |\psi_n\rangle_B \langle \psi_n |$ can be purified in ${\cal H}_{AB}$ with an entangled state 
$|\Psi\rangle_{AB} = \sum_n \sqrt{w_n}|\phi_n \rangle_A |\psi_n \rangle_B$, and $\rho_B = \mbox{Tr}_A[|\Psi\rangle_{AB} \langle \Psi |]$. 
Similarly, we have $\rho_E = \mbox{Tr}_R[| A \rangle_{ER} \langle A |]$ for some pure state $| A \rangle_{ER} \in {\cal H}_{ER}$.
In an enlarged Hilbert space 
the process of thermalization can be represented as a unitary evolution of the combined system $ABER$. This is given by
 $\rho_{ABER} = |\Psi\rangle_{AB} \langle \Psi | \otimes |A \rangle_{ER} \langle A | \rightarrow  \rho_{ABER}' = 
( {I}_A \otimes U_{BER}) |\Psi\rangle_{AB} \langle \Psi | \otimes |A \rangle_{ER} \langle A | ( {I}_A \otimes U_{BER})^{\dagger}$, so
that $\mbox{Tr}_{AER}[\rho_{ABER}'] = \rho_B' = \exp(-\beta H)/Z$, where 
$Z = \mbox{Tr}[\exp(-\beta H)]$, and it is a thermal equilibrium state described by a canonical distribution. 
Let $\{|l \rangle\}$ be the eigenstate of the equilibrium density operator $\rho_B' = \rho_{eq}$ so that 
$\rho_{eq} = \frac{1}{\sqrt Z} \sum_l \exp(-\beta E_l/2) |l \rangle \langle l| $. 
We can use the language of quantum operation to describe the process of thermalization. The quantum operation elements for thermalization 
are given by $F_{ln} = 
\frac{1}{\sqrt Z} \exp(-\beta E_l/2) |l \rangle \langle \psi_n|$ and they satisfy $\sum_{ln} F_{ln}^{\dagger} F_{ln} = {I}$.
This channel we call `thermalizing channel' as it can transform any arbitrary density operator to a thermal state. For any 
$\rho_B$, we see that the thermalizing channel transforms $\rho_B$ to $\rho_{eq}$ as 
\begin{eqnarray}
\rho_B \rightarrow \sum_{ln}  F_{ln} \rho_B F_{ln}^{\dagger}  =  
\frac{1}{ Z} \sum_l e^{-\beta E_l/2} |l \rangle \langle l|.
\end{eqnarray}
This quantum operation has the unitary representation as 
\begin{eqnarray}
|\Psi\rangle_{AB} |A \rangle_{ER} & \rightarrow & \sum_k \sum_{ln}\sqrt{w_k} |\phi_k \rangle_A  F_{ln} |\psi_k\rangle_B |A_{ln} \rangle_{ER} \nonumber\\
& = & \frac{1}{\sqrt Z} \sum_{kl} \sqrt{w_k} e^{-\beta E_l/2}  |\phi_k \rangle_A |l \rangle_B |A_{lk} \rangle_{ER}, \nonumber\\
\end{eqnarray}
where $|A_{lk} \rangle$ are the orthonormal basis for the ancilla Hilbert space ${\cal H}_{ER}$. From the above transformation one can see that $\rho_B'$ is a 
canonical thermal state. The thermalizing channel completely decorelates the quantumness of the original state $\rho_{AB}$. 
Again, we can check that the cost of erasing quantum correlation respects our bound $S(\rho_B) \le \Delta S_T$.

Before we conclude, we observe the following. When the subsystem $B$ of a quantum system $S=AB$ interacts with the environment $E$, the total correlation in 
$AB$ as quantified by the mutual information $I(\rho_{AB})$ decreases. Does it get compensated somewhere else?  Indeed it is. We can show that the reduction 
in the total correlation between $A|B$ cannot exceed the increase in the total correlation between $AB|E$. More precisely, one can prove
that $\Delta I_{A|B} \le \Delta I_{AB|E}$, where  $\Delta I_{A|B} = I(\rho_{AB}) - I(\rho_{AB}')$ and 
$\Delta I_{AB|E} = \Delta I_{S|E} = I(\rho_{SE}') - I(\rho_{SE})$. Note that mutual cannot increase under quantum operation, hence 
$\Delta I_{A|B} \ge 0$ \cite{partovi1}. Since $\rho_{SE} = \rho_S \otimes \rho_E$, quantum mutual information $I(\rho_{SE}) =0$. 
Now, $I(\rho_{SE}') = S(\rho_{S}') + S(\rho_{E}') - S(\rho_{SE}')=  \Delta S_{AB} + \Delta S_E = \Delta S_T$. Since we 
have $\Delta I_{A|B} \le \Delta S_T$, hence, the proof.

In conclusion, we have shown that any physical process that acts locally on one of the subsystem and leads to change in 
the quantum correlation, then that must be accompanied by the entropy generation of the combined system and environment. If the 
physical process leads to complete erasure of quantum correlation then the cost of erasing quantum correlation cannot be smaller than the lost 
quantum correlation. Thus, {\em there is no physical process whose sole action is to erase quantum correlation without increasing 
entropy of the combined system and the environment}. 
 In addition, we have shown that the amount of quantum correlation created is at least equal to the total negentropy of the system and the environment.
Thus, erasure and creation of quantum correlation both have thermodynamic cost. We have shown that our new bound leads to the standard Landaue erasure as
well as a generalized version of the Landauer erasure in the presence of quantum memory. Moreover, our new inequality can be applied to  
bleaching of quantum information, thermalization and many other physical processes that lead to loss of quantumness. Since quantum discord 
is related to other measure of quantum correlation like work deficit, our result suggests that erasure of other measures of 
quantum correlations can have physical cost. We hope that our results provide a foundational basis 
for the physical nature of quantum correlation independent of any particular protocol and can have deep implication in understanding the 
nature of quantum correlations.


\vskip .5cm

\noindent
{\bf Acknowledgement} AKP thanks A. K. Rajagopal, A. Sen De and U. Sen for useful discussions.

\end{document}